\documentclass[french]{gretsi}

\usepackage[utf8]{inputenc}

\usepackage[T1]{fontenc}

\usepackage{amsmath, amssymb, amsfonts}
\usepackage{array}

\begin{document}


\titre{PTSD-MDNN~: Fusion tardive de réseaux de neurones profonds multimodaux pour la détection du trouble de stress post-traumatique}

\auteurs{
  \auteur{Long}{Nguyen-Phuoc}{long.nguyen-phuoc@emse.fr}{1,2}
  \auteur{Renald}{Gaboriau}{renald.gaboriau@mjinnov.com}{2}
  \auteur{Dimitri}{Delacroix}{dimitri.delacroix@mjinnov.com}{2}
  \auteur{Laurent}{Navarro}{navarro@emse.fr}{1}
}

\affils{
  \affil{1}{Mines Saint-Étienne, University of Lyon, University Jean Monnet, Inserm, U 1059 Sainbiose, Centre CIS, 42023 Saint-Étienne, France
  }
  \affil{2}{MJ Lab, MJ INNOV, 42000 Saint-Etienne, France 
  }
}


\resume{Afin de proposer un moyen plus objectif et plus rapide de diagnostiquer le trouble de stress post-traumatique (TSPT), nous présentons \verb!PTSD-MDNN! qui fusionne deux réseaux de neurones convolutifs unimodaux et qui donne un faible taux d'erreurs de détection. En ne prenant que des vidéos et des audios comme entrées, le modèle pourrait être utilisé dans la configuration de séances de téléconsultation, dans l'optimisation des parcours patients ou encore pour l'interaction humain-robot.}

\abstract{In order to provide a more objective and quicker way to diagnose post-traumatic stress disorder (PTSD), we present \verb!PTSD-MDNN! which merges two unimodal convolutional neural networks and which gives low detection error rate. By taking only videos and audios as inputs, the model could be used in the configuration of teleconsultation sessions, in the optimization of patient journeys or for human-robot interaction.}

\maketitle


\section{Introduction}

\subsection{Contexte Général}
Tout individu, au cours de sa vie, peut rencontrer des situations potentiellement traumatogènes. ~\cite{american_psychiatric_association_dsm-5_2015} définit comme traumatogène toute situation qui implique « une mort effective, une menace de mort, une blessure grave ou des violences sexuelles ». En France, le TSPT toucherait entre 1 et 2\% de la population. Les symptômes du TSPT causent des problèmes importants dans les situations sociales ou professionnelles.

Traditionnellement, le TSPT a été diagnostiqué par des professionnels de la santé impliquant  des questionnaires. La collecte d'informations par le biais d'un questionnaire auto-déclaratif a des limites car souvent biaisés~\cite{banerjee_deep_2019}~: (1) les distorsions de la mémoire et de la perception de soi des patients rendent également le diagnostic difficile, et (2) les patients sont souvent gênés d'être diagnostiqués et ne veulent pas visiter les cliniques pour le diagnostic. Finalement, peu de mesures objectives ou qualitatives sont disponibles pour aider les cliniciens à diagnostiquer ce trouble. Un exemple d'un tel entretien est le Clinician-Administered PTSD Scale (CAPS) ou encore le PCL-5~\cite{american_psychiatric_association_dsm-5_2015}.

La récente pandémie de SRAS-CoV-2 peut être considérée comme un événement traumatique mondial qui a fait émerger: (1) un fort impact sur la santé mental, (2) la réalisation de nombreux soins médicaux transformée en distanciel. Par conséquent, considérant le biais des modes de collecte de données auto-déclaratifs et le changement radical induit par les consultations en distanciel, il est nécessaire de trouver un moyen plus objectif et rapide pour diagnostiquer le TSPT.

\subsection{Contexte Théorique}
Nous distinguons deux catégories de modèles d'intelligence artificielle : les modèles de pronostic et les modèles de diagnostic du TSPT. Notre papier se concentre sur le diagnostic, c'est-à-dire la détection de l'état actuel des patients. Une grande majorité des études sur le diagnostic utilisent des techniques d'apprentissage automatique supervisé sur des données structurées. Par exemple,~\cite{rangaprakash_compromised_2017,breen_modelling_2019} ont tous appliqué des algorithmes de Machine à Vecteurs de Support (SVM) sur des questionnaires. Toujours sur ces données auto-déclaratives et tabulaires,~\cite{omurca_alternative_2015} ont utilisé l'optimisation minimale séquentielle, le perceptron multicouches et la classification naïve bayésienne. Concernant les données biométriques, il semble que les modifications de la variabilité de la fréquence cardiaque sont significativement associées au TSPT~\cite{hauschildt_heart_2011}. Enfin, la conductance cutanée peut être utilisée comme outil de diagnostic~\cite{hinrichs_mobile_2017} car elle semble en particulier corrélée à son intensité.

À l'image des données non structurées qui alimentent la prochaine génération des modèles d'IA, de nombreuses études de détection du TSPT ont profité de ces avancées pour puiser l'information dans différentes sources disponibles dans le cadre clinique ou quotidien des patients. Parmi les moyens d'acquisition et de restitution d'imagerie médicale, les études d'imagerie par résonance magnétique (IRM) structurelle~\cite{salminen_adaptive_2019} et fonctionnelle~\cite{rangaprakash_compromised_2017} ont permis des progrès considérables dans la compréhension des mécanismes neuronaux sous-jacents au TSPT. Ainsi couplées avec l'agorithme SVM, elles permettent de détecter le TSPT avec de bon résultats.

\begin{figure*}[htb]
  \begin{center}
    \includegraphics[width=2\columnwidth]{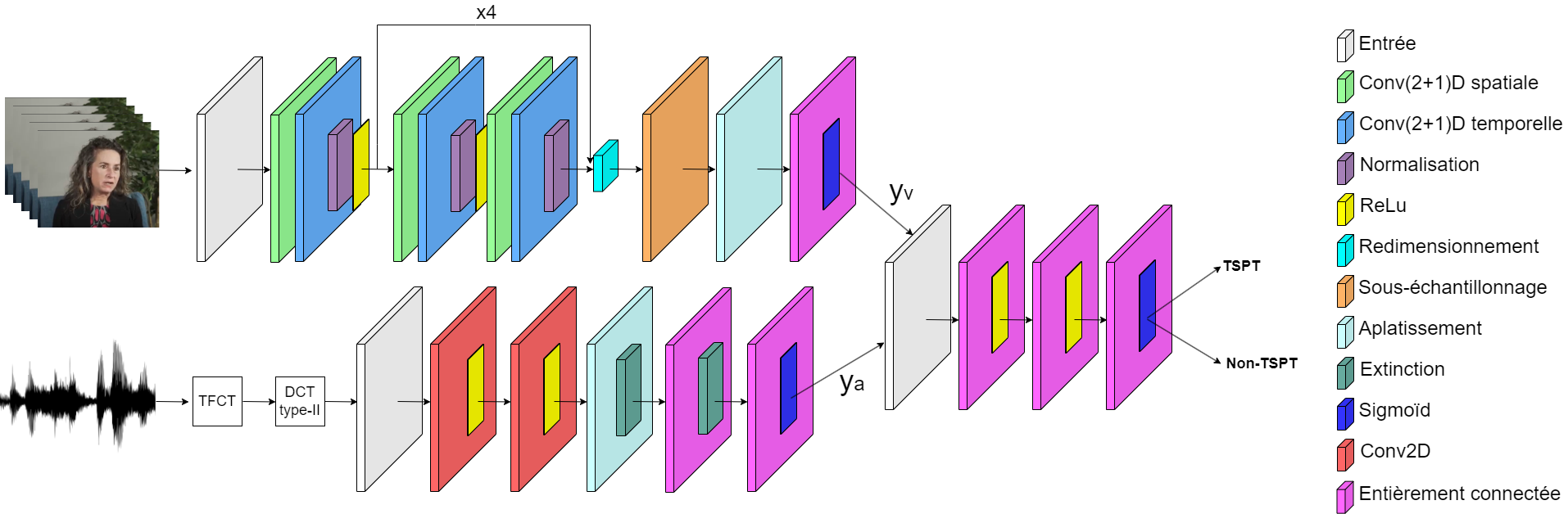}
  \end{center}
  \caption{L'architecture de PTSD-MDNN}
  \label{architecture}
\end{figure*}

D'autres études ont cherché des alternatives aux données médicales traditionnelles. Le diagnostic des patients atteints de TSPT par l'analyse des signaux vocaux a été étudié depuis ces dernières années~\cite{suneetha_survey_2022}. Des approches de text mining ou de traitement du langage naturel ont été également utilisées~\cite{banerjee_deep_2019,sawalha_detecting_2022}. Même si la télémédecine via vidéo conférence  peut réduire le délai de la prise en charge et être aussi efficace que le traitement en personne~\cite{lindsay_implementation_2015}, seul~\cite{sawadogo_ptsd_2022} se concentre sur ce type de données audiovisuelles en utilisant différentes architectures de réseaux de neurones par types de modalités de données. Actuellement, la question de la fusion de ces différentes modalités audio-vidéo pour la détection du TSPT reste entière.

\subsection{Motivation}
Malgré la variété des études citées ci-dessus, le diagnostic du TSPT nécessite des capteurs ou dispositifs médicaux à caractère invasif dont la disponibilité n'est pas garantie, notamment en temps de crise. Nous proposons, dans ce papier, un réseau de neurones profond multimodal pour la détection automatique de TSPT (\verb!PTSD-MDNN:! \verb!Post Traumatic Stress Disorder! \verb! - Multimodal Deep Neural Network!) en utilisant comme entrées de simples vidéos et audios des patients en situation réelle. Nous montrons qu'en plus d'être adapté à ces données audiovisuelles facilement collectées, notre modèle obtient de meilleurs résultats en fusionnant ces deux modalités.


\section{Méthode Proposée} \label{methode}
Nous présentons un aperçu de \verb!PTSD-MDNN! dans la Fig~\ref{architecture}. Le modèle général se compose de deux sous-modèles qui prennent chacun en entrée une modalité différente. Le vecteur sortant de la dernière couche du classement vidéo \(y_v\) est concatené avec le vecteur sortant de la dernière couche du classement audio \(y_a\) pour former une matrice \(M_{v,a}\) . Après cette fusion tardive de modalités, la matrice \(M_{v,a}\) est injectée dans un dernier réseau de neurones à deux couches afin de détecter le TSPT.

\subsection{Classement Vidéo}
\begin{figure}[htb]
  \begin{center}
    \includegraphics[width=1\columnwidth]{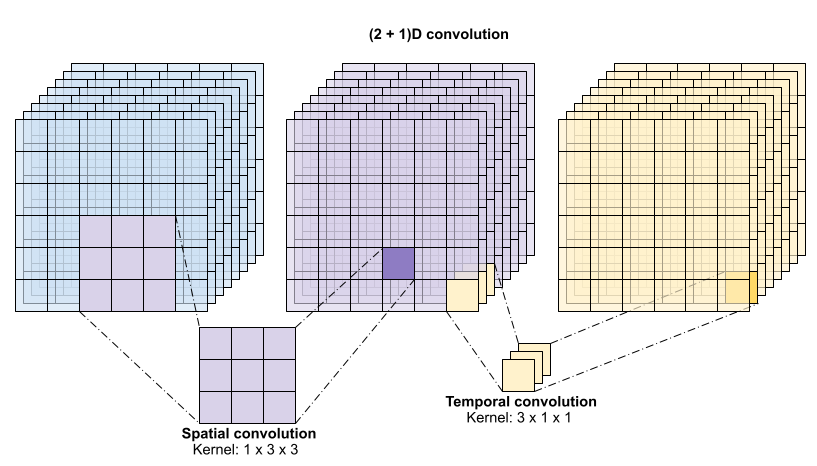}
  \end{center}
  \caption{Les convolutions spatiales et temporelles factorisées d'une convolution (2+1)D avec une taille de noyau \((3 \times 3 \times 3)\) nécessitent des matrices de poids de taille \((9 \times canaux^2) + (3 \times canaux^2)\). Ceci est moins de la moitié de celles nécessaires pour la convolution 3D complète \((27 \times canaux^2)\) }
  \label{2plus1}
\end{figure}

Le sous-modèle de classification vidéo utilise un réseau de neurones convolutif (2+1)D~\cite{tran_closer_2018} avec des connexions résiduelles à 18 couches de profondeur (ResNet18). La convolution (2+1)D permet la décomposition des dimensions spatiale et temporelle, créant ainsi deux étapes distinctes. Un avantage de cette approche est que la factorisation des convolutions en dimensions spatiales et temporelles permet de réduire le nombre de paramètres par rapport à la convolution 3D complète. La convolution spatiale prend les données sous la forme \((1, largeur, hauteur)\), tandis que la convolution temporelle prend les données sous la forme \((temps, 1, 1)\) comme illustré dans la Fig~\ref{2plus1}. Le redimensionnement de la vidéo est nécessaire pour: (1) effectuer un sous-échantillonnage des données, (2) examiner des parties spécifiques des images, (3) réduire la dimensionnalité pour un traitement plus rapide.

\subsection{Classement Audio}
Nous avons adapté le modèle~\cite{kahl_overview_2022} basé sur une transformation de Fourier à court terme (TFCT) avec des fenêtres plus longues (64 ms) chevauchées à 75\%, ce qui donne une meilleure résolution en fréquence pour la voix humaine. Nous avons ensuite converti les fréquences du spectrogramme en échelle logarithmique de Mel qui se rapproche de la perception humaine du son. Finalement, la transformée en cosinus discrète de type II (DCT) donne les coefficients Mel-Frequency Cepstral (MFCC) par 80 filtres triangulaires créés pour couvrir la plage de fréquences Mel. Nous sélectionnons uniquement 13 premiers qui sont utiles à la reconnaissance de la parole~\cite{breebaart_features_2004}.

Le classement audio est entraîné à partir de ces MFCC en utilisant un réseau de neurones convolutif illustré dans la Fig~\ref{architecture}. Le modèle commence par deux couches de convolution 2D avec 16 filtres chacune, une taille de noyau de (3,3) et une fonction d'activation ReLu. Ces couches convolutives 2D sont utilisées pour extraire des caractéristiques importantes des MFCC d'entrée, qui sont des matrices de taille (778, 13, 1). Ensuite, les couches d'Aplatissement (Flatten) et d'Extinction (Dropout) s'enchainent pour respectivement convertir la sortie de la dernière couche convolutive en un vecteur à une dimension et pour désactiver aléatoirement certains neurones afin de contourner le surapprentissage. Deux couches entièrement connectées sont ensuite combinées via une fonction Sigmoïd, ce qui permet d'obtenir les probabilités des classes à prédire.

\subsection{Fusion Tardive Des Modalités}
Nous avons choisi pour notre modèle une fusion tardive, dite << fusion orientée décisions >>. En effet,~\cite{wang_what_2020} a montré que~: (1) un meilleur réseau unimodal peut surpasser un réseau multimodal contre le problème de surapprentissage; (2) différentes modalités se surajustent et se généralisent à des rythmes différents, donc les entraîner conjointement avec une seule stratégie d'optimisation n'est pas optimal. Nous avons imaginé un mécanisme de << correction d'erreur >> qui fusionne des prédictions provenant de deux réseaux unimodaux qui sont entrainés séparément. La fonction de perte que nous avons utilisé pour ce réseau de fusion tardive est la même que pour les sous-modèles: Perte d’entropie croisée binaire définie comme la formule~\ref{formule}. Cette formule suppose que les \(p_i\) sont des probabilités et les \(y_i\) sont des labels (0;1).

\begin{equation}
  \label{formule}
  L = -\frac{1}{N}\sum_{i=1}^{2}y_ilog(p_i) = -\frac{1}{N}[y_1log(p_1)+y_2log(p_2)]
\end{equation}


\section{Évaluation} \label{Evaluation}

\subsection{Base De Données}
En général, il est difficile de collecter des données de haute qualité auprès de personnes qui présentent des symptômes de TSPT. Il peut y avoir aussi des considérations éthiques qui limitent la collecte et l'utilisation de données en milieu naturel. Cela peut être particulièrement difficile dans le contexte d'un trouble sensible comme le TSPT, où les participants peuvent être réticents à divulguer des informations personnelles.

Par conséquent, seules quatre bases de données non structurées pour la détection du TSPT existent: eDAIC-WOZ~\cite{gratch_distress_2014}, FEMH~\cite{islam_transfer_2018}, Aurora~\cite{rathlev_correction_2020} et PTSD in-the-wild~\cite{sawadogo_ptsd_2022}. Parmi elles, seules les bases eDAIC-WOZ et PTSD in-the-wild disposent à la fois de modalités audio et vidéo. Nous avons choisi d'appliquer notre modèle \verb!PTSD-MDNN! sur la base PTSD in-the-wild pour son caractère réel en milieu naturel. La base de données PTSD in-the-wild (EULA) contient 634 vidéos équilibrées~: 317 vidéos de sujets avec TSPT et 317 vidéos de sujets témoins sains avec aucun symptôme de TSPT.

\subsection{Résultats}
Nous nous intéressons à l'évaluation d'une classification binaire avec deux classes~: TSPT (positif) et Non-TSPT (négatif) avec différentes métriques de classification populaires~: l'accuracy, la précision, le rappel. Nous avons suivi le processus train/validation/test proposé par \cite{sawadogo_ptsd_2022} (80\%/10\%/10\%) pour entraîner (\verb!PTSD-MDNN!) sur une carte GPU NVIDIA A100 SXM 40Go avec taille de batch de 8, un taux d'apprentissage de 0.001, et un optimiseur Adam pour 50 époques. Ces paramètres sont optimaux pour éviter le surapprentissage lié à la taille des données. De plus, nous avons utilisé différentes méthodes de régularisation pour le classement audio~(Table~\ref{results}).

\begin{table}[htb]
    \caption{\label{results}Résultats du classement sur les données test}
    \small
    \begin{center}
    \setlength{\tabcolsep}{2.5pt}
    \begin{tabular}{l*{6}{c}}
        \toprule
        Modalité   & Régularisation &  Accuracy &   Précision &   Rappel \\
        \midrule
        Vidéo & NA & 0,89 & 0,84 &  0,84 \\
        Audio & NA & 0,72 & 0,68 &  0,81\\
        Audio & L1 &  0,73 &  0,67 &  0,90 \\
        Audio & L2 &  0,75 & 0,72 & 0,81 \\
        Vidéo + Audio & L1 & 0,89 &  0,90 & 0,87 \\
        Vidéo + Audio & L2 & 0,92 & 0,88 & 0,97 \\
        \bottomrule
    \end{tabular}
    \end{center}
\end{table}

\subsection{Discussion}
Le meilleur des modèles unimodaux est le classement vidéo avec une accuracy de 0,89. En revanche, le classement audio de base ne donne pas de très bons résultats même si les régularisations L1 et L2 l'améliorent respectivement à 0,73 et 0,75. Notre approche de fusion tardive des modalités apporte de réelles améliorations par rapport aux classements unimodaux car \verb!PTSD-MDNN!  donne la meilleure accuracy (0,92), avec une régularisation L2, et le meilleur rappel (0,97). Le principal avantage d'une fusion tardive est la prise en charge des différentes modalités non alignées ainsi nous ne dépendons pas de l'interopérabilité des capteurs. De plus, l'entrainement indépendant des deux sous-modèles permet de gagner du temps en effectuant des tâches en parallèle. Enfin, cette fusion permet d'apporter la flexibilité pour le choix des sous-modèles adaptés à chaque modalité. Il est à noter que la différence de taille des fichiers de la base  PTSD in-the-wild (le plus court~: 0~min~35~s et le plus long~: 44~min~40~s) peut créer des difficultés pour l'extraction des variables en entrée du modèle de convolution.


\section{Conclusion}
Nous proposons \verb!PTSD-MDNN!, un modèle qui fusionne tardivement des modalités audio et vidéo pour détecter le TSPT. Grâce à un mécanisme de correction d'erreur, notre modèle surpasse les modèles unimodaux. En plus d’être non invasif, \verb!PTSD-MDNN! traite les informations sensibles sur les patients à très bas niveau (pixel, MFCC), ce qui permet de garder une certaine confidentialité pour les patients par rapport aux approches type NLP où les paroles sont transcrites.

Ce travail ouvre une multitude de travaux futurs. Premièrement, nous avons l'intention d'extraire des variables de haut niveau à partir d'aspects comportementaux subtils, tels que les mouvements du corps, les expressions faciales pour la vision ainsi que la prosodie et la parole pour l'audio. Deuxièment, d'autres directions concernent la fusion des modalités à travers le mécanisme de l'attention inter-modalité.

\bibliography{biblio}

\nocite{*} 


\end{document}